\documentclass[pra,reprint,superscriptaddress]{revtex4-1}
\usepackage{hyperref}
\usepackage{amsmath}
\usepackage{textcomp}
\usepackage{siunitx}
\usepackage{graphicx}
\usepackage{multirow}
\usepackage{dcolumn}
\newcommand{\SIadj}[2]{\SI[number-unit-product={\text{-}}]{#1}{#2}}
\let\mc\multicolumn

\begin{document}
\title{Precision spectra of $A\, ^2\Sigma^+,v'=0 \leftarrow X\, ^2\Pi_{3/2},v''=0,J''=3/2$ transitions in $^{16}$OH and $^{16}$OD}
\date{\today}
\author{Arthur Fast}
\affiliation{Max Planck Institute for Biophysical Chemistry, Am Fassberg 11, 37077 G\"ottingen, Germany}
\author{John E. Furneaux}
\affiliation{Max Planck Institute for Biophysical Chemistry, Am Fassberg 11, 37077 G\"ottingen, Germany}
\affiliation{Homer L. Dodge Department of Physics and Astronomy, University of Oklahoma, 440 W.\ Brooks St., Norman, OK 73019, USA}
\author{Samuel A. Meek}
\email{samuel.meek@mpibpc.mpg.de}
\affiliation{Max Planck Institute for Biophysical Chemistry, Am Fassberg 11, 37077 G\"ottingen, Germany}

\begin{abstract}
We report absolute optical frequencies of electronic transitions from the $X\, ^2\Pi_{3/2},v''=0,J''=3/2$ rovibronic ground state to the 12 lowest levels of the $A\, ^2\Sigma^+,v'=0$ vibronic state in $^{16}$OH, as well as to the 16 lowest levels of the same vibronic state in $^{16}$OD.   
The absolute frequencies of these transitions have been determined with a relative uncertainty of a few parts in $10^{11}$, representing a $\sim$1000-fold improvement over previous measurements.
To reach this level of precision, an optical frequency comb has been used to transfer the stability of a narrow-linewidth I$_2$-stabilized reference laser onto the \SIadj{308}{nm} spectroscopy laser.
The comb is also used to compare the optical frequency of the spectroscopy laser to an atomic clock reference, providing absolute accuracy.
Measurements have been carried out on OH/OD molecules in a highly-collimated molecular beam, reducing possible pressure shifts and minimizing Doppler broadening.
Systematic shifts due to retroreflection quality, the Zeeman effect, and the ac Stark effect have been considered during the analysis of the measured spectra; particularly in the case of the OD isotopologue, these effects can result in shifts of the fitted line positions of as much as \SI{300}{kHz}. 
The transition frequencies extracted in the analysis were also used to determine spectroscopic constants for the $A\, ^2\Sigma^+,v'=0$ vibronic state.  
The constants fitted in this work differ significantly from those reported in previous works that measured the $A - X$ transitions, resulting in typical deviations of the predicted optical transition frequencies of \SI{\sim 150}{MHz}, but they generally agree quite well with the constants determined using hyperfine-resolved measurements of splittings within the $A$ state.
\end{abstract}

\maketitle

The hydroxyl radical, OH, is a prototypical open-shell diatomic molecule that is important in a variety of fields, including atmospheric chemistry \cite{stone12}, interstellar chemistry \cite{green81}, crossed-beam molecular collision studies \cite{kirste12}, and Stark deceleration \cite{meerakker06}.
In laboratory studies, OH is commonly detected with rotational state selectivity by measuring laser-induced fluorescence from ultraviolet $A\, ^2\Sigma^+ - X\, ^2\Pi$ transitions.
More-recently, a sensitive detection scheme based on 1+1' resonance-enhanced multiphoton ionization (REMPI) was demonstrated which also makes use of $A \leftarrow X$ excitation as a first step \cite{beames11}.
Previous studies have determined the absolute frequencies of the $A-X$ transitions with an uncertainty of approximately \SI{0.005}{cm^{-1}} (\SI{150}{MHz}) \cite{coxon75,coxon80,stark94}.
This level of accuracy is quite sufficient for excitation with commonly-used frequency-doubled pulsed dye lasers, which typically have a bandwidth on the order of \SI{0.1}{cm^{-1}}, but for driving the transitions with a continuous-wave (cw) laser with a linewidth on the order of \SI{1}{MHz} or less, the transition frequencies must be known much more exactly.

In this work, we present high-precision measurements of the $A\,^2\Sigma^+,v'=0 \leftarrow X\, ^2\Pi_{3/2},v''=0,J''=3/2$ transitions in $^{16}$OH and $^{16}$OD.
Using a frequency-doubled cw dye laser which is stabilized and monitored with the help of an optical frequency comb, we have measured transitions to the 12 lowest levels of the $A\,^2\Sigma^+,v'=0$ vibronic state of $^{16}$OH and to the 16 lowest $A$ levels in $^{16}$OD with an experimental uncertainty of a few tens of kHz, or a few parts in $10^{11}$ relative uncertainty. 
These measurements have enabled us to determine spectroscopic constants such as the $A\,^2\Sigma^+,v'=0$ band origin and the rotational constant $B$ with orders of magnitude higher precision than previously possible.

\section{Experimental setup}
The experimental setup can be roughly divided into two major components: a precision laser system, which is used in this work to generate a frequency-stable, narrow-linewidth cw beam with a wavelength near \SI{308}{nm} whose absolute frequency is known relative to atomic clock references, and a molecular beam apparatus for producing packets of rotationally-cold OH radicals in vacuum and detecting the fluorescence induced by the spectroscopy laser.  

\subsection{Precision laser system}
\begin{figure}
\includegraphics{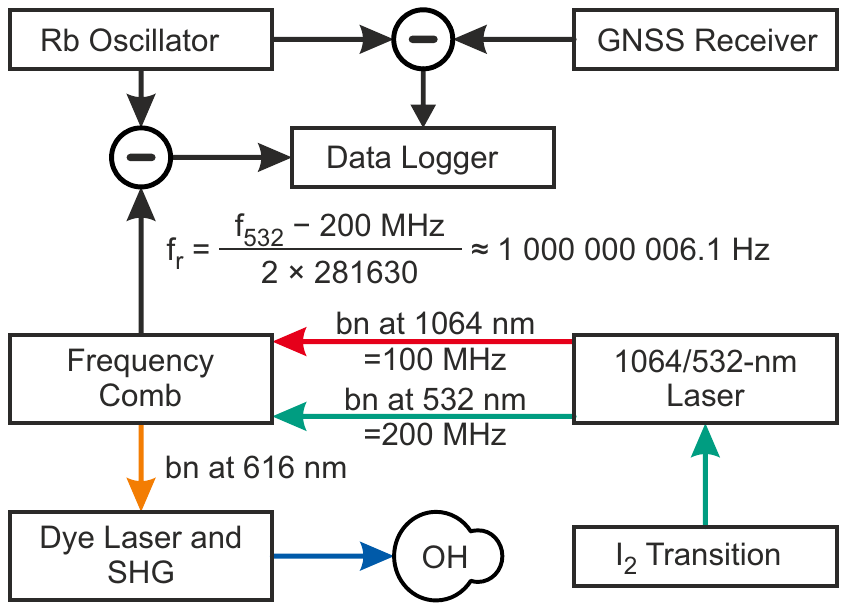}
\caption{Schematic overview of the precision laser system used to produce a tunable, narrow-linewidth \SIadj{308}{nm} cw beam for spectroscopy of the low-lying $A-X$ transitions in OH/OD.  The optical frequency of the UV source is stabilized to an I$_2$-referenced \SIadj{1064/532}{nm} laser using an optical frequency comb as a transfer oscillator.  The comb is also used to monitor the laser's optical frequency relative to a local rubidium oscillator and a global navigation satellite system (GNSS) receiver, providing a record of its absolute optical frequency during the measurement.}
\label{fig-setup-laser}
\end{figure}
A schematic diagram of the precision laser system is shown in figure \ref{fig-setup-laser}.
The foundation of the laser system is a short-term frequency reference based on a \SIadj{1064}{nm} cw neodymium-doped yttrium aluminum garnet (Nd:YAG) laser (Coherent Mephisto 1000 NE), part of which is frequency doubled to \SI{532}{nm} using a periodically-poled lithium niobate (PPLN) waveguide (NTT Electronics).
Using a similar apparatus to that described by D\"orringshof et al.\ \cite{doeringshoff10}, the laser's optical frequency is stabilized to the $a_{10}$ component of the R(56) 32--0 transition in molecular iodine (I$_2$) using saturated absorption spectroscopy.
The optical frequency of the stabilized \SIadj{532}{nm} laser is approximately $f_{532} = \SI{563260203.42}{MHz}$ and has remained stable to within \SI{50}{kHz} over several months.
On shorter timescales (\SIrange[range-units=single]{100}{10000}{s}), comparisons of this reference to a rubidium oscillator, using the method described in the next paragraph, show an Allan deviation of at most a few parts in $10^{13}$, which is likely limited by the rubidium reference.
The system constructed by D\"orringshof et al.\ was shown to have an Allan deviation of approximately $10^{-14}$ at these timescales.

An optical frequency comb is then stabilized to this optical frequency reference.
The comb is based on a Ti:Sapphire femtosecond oscillator with a \SIadj{\sim 1}{GHz} repetition rate (Taccor-6 from Laser Quantum) which is broadened to cover the spectral region from \SIrange[range-units=single]{1100}{500}{nm} using a photonic crystal fiber module (NKT Femtowhite 800).  
The two degrees of freedom that determine the absolute frequency of every comb mode (commonly defined in terms of the repetition rate $f_r$ and the carrier-envelope offset frequency $f_0$ \cite{holzwarth00}) are stabilized to the iodine frequency reference using the combination of an optical beatnote between the comb and the \SIadj{1064}{nm} beam and a beatnote between the comb and the \SIadj{532}{nm} beam.
Specifically, a pair of phase-locked loops ensure that one mode of the comb has an optical frequency exactly \SI{100}{MHz} lower than the \SIadj{1064}{nm} beam and that another comb mode is exactly \SI{200}{MHz} lower in frequency than the \SIadj{532}{nm} beam.  
These constraints force $f_0$ to be zero and result in a direct link between $f_r$ and $f_{532}$ given by
\begin{equation}
f_r = \frac{f_{532} - \SI{200}{MHz}}{2 n}\rm{,}
\label{eqn-frep}
\end{equation}
where $n$ is an integer describing the number of comb modes between the \SIadj{1064}{nm} and \SIadj{532}{nm} frequencies.  
For the measurements shown in this paper, $n$ is either \num{281630} (as shown in figure \ref{fig-setup-laser}) or \num{281631}, resulting in a repetition rate of \SI{1000000006.1}{Hz} or \SI{999996455.3}{Hz}, respectively.
While the exact value of $f_r$ can vary over time, due to slight drifts of the iodine-locked optical reference, absolute accuracy is nonetheless achieved by recording the comb's repetition rate on deadtime-free frequency counter.  
The counter is referenced to a rubidium oscillator (PRS10 from Stanford Research Systems) which is stabilized against long-term drifts using a pulse-per-second (PPS) signal from a global navigation satellite system (GNSS) receiver (PolaRx4TR PRO from Septentrio).
The GNSS receiver also records data that can be used to reconstruct the phase error of the rubidium oscillator relative to GNSS time, enabling further post-correction of the frequency offset.
For the precision of the current measurements, however, this extra step was not found to be necessary: when averaged over \SIadj{1000}{second} intervals, typical for a single measurement scan, the relative root mean square deviation between the frequency given by GNSS time and that produced by the rubidium oscillator is found to be less than $1.5 \cdot 10^{-12}$.
Radio-frequency (rf) reference signals for stabilizing the various beatnotes are also derived from the rubidium oscillator, so any beatnote frequencies specified in this work (such as the ``\SI{200}{MHz}'' in equation \ref{eqn-frep}) are defined relative to this reference.  

The \SIadj{308}{nm} spectroscopy laser is based on a tunable cw dye laser operating at 616 nm (Matisse 2 DR from Sirah GmbH) which is subsequently converted to 308 nm through second-harmonic generation (SHG) using a beta barium borate (BBO) crystal in an enhancement cavity (WaveTrain 2 from Sirah).  
The optical frequency of the dye laser is compared to the frequency comb using an optical beat note and stabilized with a tunable frequency offset (specified using a computer-controlled rf generator) to the nearest comb mode using a feedback loop that controls piezo mirrors on the dye laser's cavity.
To scan over each transition, the frequency of the rf generator is adjusted in \SI{100}{kHz} steps over a typical span of between \SIrange[range-phrase={ and },range-units=single]{16}{25}{MHz}.  
Unfortunately, it is not possible to stabilize the dye laser's optical frequency at every offset frequency between 0 and $f_r/2$: if the beatnote frequency is too low, there is ambiguity as to whether the laser's frequency is higher or lower than the comb mode, and if it is too high, there is ambiguity between the beatnote with comb mode below and with the comb mode above the laser's frequency.
While the full range of beatnote frequencies could theoretically, with our \SIadj{1}{GHz} comb, span from \SIrange[range-units=single]{0}{\sim 500}{MHz}, the feedback loops can only operate properly when the offset frequency is between \SIrange[range-phrase={ and }]{50}{450}{MHz}.  
If the comb were only stabilized at the repetition rate corresponding to $n = \num{281630}$ (from equation \ref{eqn-frep}), measuring certain transitions would require scans that, at least partially, overlap with the dead zones.
Fortunately, all of the transitions that fall into dead zones for $n = \num{281630}$ end up with more favorable offset frequencies for $n = \num{281631}$, so all transitions presented in this work could be measured using one of these two comb repetition rates.  

\subsection{Molecular beam}\label{sec-setup-mol}
\begin{figure}
\includegraphics{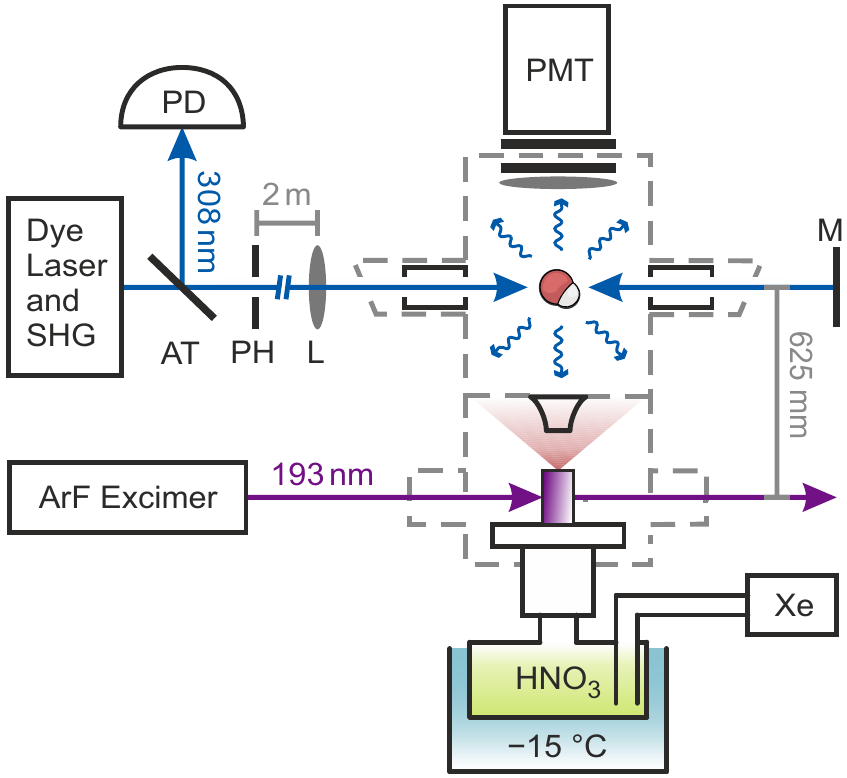}
\caption{Schematic overview of the OH molecular beam apparatus.  OH/OD molecules are produced through photodissociation of (deuterated) nitric acid with an ArF excimer laser in a xenon-seeded pulsed supersonic molecular beam.  Approximately \SI{625}{mm} downstream, the molecules are excited by the \SIadj{308}{nm} spectroscopy laser, emitting fluorescence which is detected by a photomultiplier tube (PMT).  Near its source, the spectroscopy laser passes through a reflective attenuator (AT) and a \SIadj{0.8}{mm} pinhole (PH), after which it travels approximately \SI{2}{m} through free space, diverted only by flat mirrors.  Immediately before entering the vacuum chamber, the laser is collimated by a \SIadj{2}{m} focal length lens (L), and after exciting the chamber on the other side is retroreflected by a flat mirror (M).  The retroreflected beam returns through the pinhole, reflects off the attenuator, and is detected on a photodiode (PD).  By maximizing the intensity of the beam reaching the photodiode, the residual angle between the counterpropagating beams is minimized.}
\label{fig-setup-mol}
\end{figure}
Figure \ref{fig-setup-mol} shows a schematic diagram of the molecular beam apparatus used in the experiment.  
The OH/OD molecules in the molecular beam are produced through the photodissocitation of nitric acid (HNO$_3$/DNO$_3$) with a \SIadj{193}{nm} ArF excimer laser.
First, a dilute mixture of nitric acid vapor in xenon is produced by passing xenon gas through reservoir containing white fuming nitric acid soaked onto glass wool.  
The reservoir is chilled to \SI{-15}{\degreeCelsius} to prevent condensation elsewhere in the gas system; this limits corrosion further downstream, and as Scharfenberg has noted \cite{scharfenberg12}, has little effect on the OH density in the final molecular beam.  
The mixture is then expanded in \SIadj{\sim 100}{\micro s} pulses through a Series 9 Parker General Valve with a \SIadj{6}{mm} long quartz capillary attached to the output of its \SIadj{1}{mm} nozzle.  
The excimer laser (GAM EX5/250-180, producing \SI{\sim 8}{mJ}, \SI{10}{ns} pulses) is weakly focused onto a \SI{1 x 2}{mm} patch near the tip of the capillary to dissociate the HNO$_3$/DNO$_3$ molecules just before the gas mixture expands into vacuum, producing a rotationally-cold sample of OH/OD molecules after the expansion.
Based on time of flight measurements, the OH/OD molecules have a longitudinal velocity of approximately \SI{340}{m/s}.

The expanding molecular beam passes through a \SIadj{4}{mm} skimmer into a second differentially-pumped chamber containing the traveling-wave Stark decelerator described in \cite{meek11a}.
This decelerator has a \SIadj{4}{mm} wide circular profile and is approximately \SI{480}{mm} long.  
For the current experiments, the decelerator electrodes are simply grounded, which results in a narrow transverse velocity spread in the molecular beam (\SI{\sim 2.5}{m/s} full width at half maximum) and ensures a negligibly-small electric field strength in the spectroscopy region at the expense of molecular density.  
After passing through the decelerator, the molecules interact with the \SIadj{308}{nm} spectroscopy laser, producing fluorescence which is collected with a \SIadj{50}{mm} fused silica lens and directed onto an on-axis photomultiplier tube (PMT, model 9829QSB from ET Enterprises).  
To reduce the intensity of the light from the photodissociation pulse that reaches the PMT, a UG5 color filter has been inserted just behind the collection lens and a UG11 filter directly in front of the PMT.
The gain of the PMT is also suppressed for a \SIadj{20}{\micro s} interval around the excimer laser pulse by switching the photocathode to a more positive potential than the first dynode.  

The \SIadj{308}{nm} laser originates from a small waist in the BBO crystal and travels approximately 2.3 meters before reaching the Brewster window at the entrance of the vacuum chamber.
Immediately before entering the chamber, the beam is collimated with a \SIadj{2}{m} focal length lens, resulting in an approximately \SI{0.8 x 1.0}{mm} wide beam in the spectroscopy region.  
Inside the chamber, a total of four \SIadj{5}{mm} circular light baffles (two before the spectroscopy region and two after) help to shield stray light from the photomultiplier.
The beam exits the vacuum chamber through a second Brewster window and is retroreflected along the same path.
Exciting the molecules using two beams exactly anti-parallel to one another helps to eliminate residual Doppler shifts due to a non-zero average velocity of the molecules along the propagation axis of the laser.  
To help ensure the retroreflected beam is as anti-parallel to the original beam as possible, a \SIadj{0.8}{mm} circular aperture is placed approximately \SI{27}{cm} after the source waist, or about \SI{2}{m} before the collimating lens.
The retroreflection is then adjusted in order to maximize the fraction of the returning beam that passes through the original aperture.
Using this technique, we estimate that the offset between the outgoing and returning beams at the aperture can be reduced to less than \SI{0.1}{mm}.
More details of the retroreflection will be discussed in section \ref{sec-sys-ref}.

\section{Measurements}
Measurements of individual transitions in OH and transition clusters in OD were carried out by scanning the frequency offset between the \SIadj{616}{nm} laser and the nearest mode of the frequency comb in \SI{100}{kHz} steps and recording the resulting fluorescence with a photomultiplier tube.
At each frequency, the analog signal at the anode of the photomultiplier (with a \SI{100}{\kilo\ohm} load to ground) was recorded on a digital oscilloscope from \SIrange[range-units=single]{1.0}{3.8}{ms} after the excimer pulse with a resolution of \SI{2}{\micro s}; measurements were averaged over 50 shots at a repetition rate of \SI{10}{Hz}.
A single scan results in a two-dimensional matrix of fluorescence intensity versus laser frequency and time delay after the excitation pulse.  
Scans over a single transition were repeated consecutively between 4 and 30 times, depending on the signal-to-noise ratio of a single measurement.  

The resulting two-dimensional matrices from the repeated scans were then averaged into a single matrix.
Since the frequencies in each scan correspond to fixed offsets from the nearest comb mode, these must first be converted to absolute frequencies using the formula
\begin{equation}
f_{abs} = 2(n f_r + f_{bn}\rm){,}
\end{equation}
where $n$ is an integer determined through frequency measurements with a wavemeter, $f_{bn}$ is the beatnote frequency, and $f_r$ is approximated as the average frequency of the comb's repetition rate over the scan.
This approximation preserves the \SIadj{200}{kHz} spacing between individual lines in the frequency scan, but due to small drifts of the reference laser frequency from one scan to the next, each scan has a slightly different frequency offset.
To account for these offsets, scans after first were shifted slightly in frequency, using linear interpolation between laser frequencies at each time delay, to match the offset in the first scan.

\begin{figure}
\includegraphics{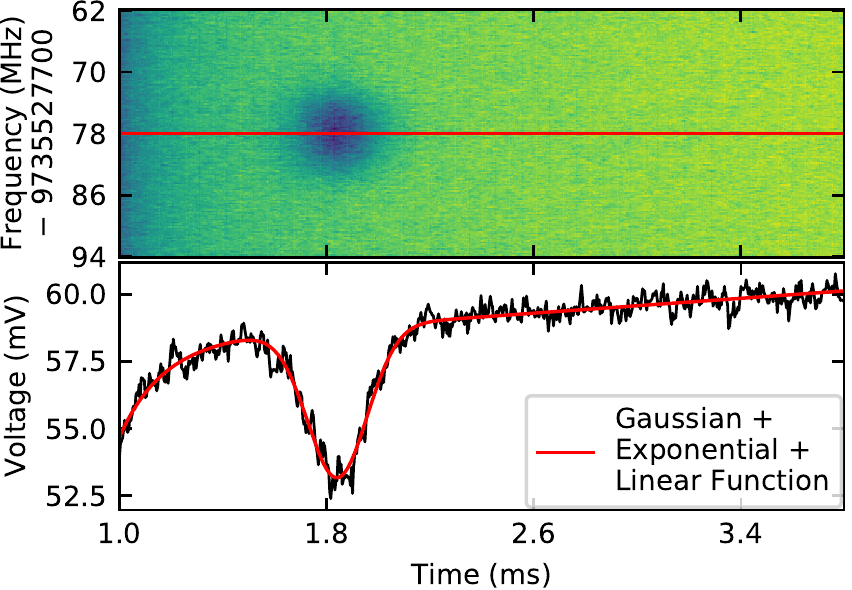}
\caption{Fluorescence intensity as a function of time delay after the excimer dissociation pulse and optical frequency of the spectroscopy laser.  The bottom half of the figure shows the fluorescence intensity as a function of time delay at a single laser frequency (indicated in the top half by a red horizontal line).  The red curve in the bottom half shows the 7-parameter fit used to determine the background intensity at this particular laser frequency.}
\label{fig-osci}
\end{figure}
An example of the resulting averaged matrix, as well as a cut-through at a single laser frequency, are shown in figure \ref{fig-osci}.
In addition to fluorescence from the packet of molecules (arriving at \SI{\sim 1.84}{ms}), there are background contributions at early times due to fluorescence resulting from the excimer pulse and a steady-state background due to scattering of the spectroscopy laser.  
To remove these background contributions, the entire time trace at each frequency is fit independently to a model containing 7 parameters: two for an exponential decay, two for a linear trend of the baseline (which likely contains one or more exponential decays with a long time constant), and three (amplitude, position, and standard deviation) for a Gaussian profile to describe the packet of molecules.  
The four components of the fit associated with background contributions are then subtracted from the trace, and the fluorescence intensity is computed by integrating over a fixed time window around the measured arrival time of the molecules.
The limits of this window are determined by summing all traces in a spectrum and performing the same 7-parameter fit as before; the region considered encompasses $\pm2$ fitted standard deviations around the fitted center position of the Gaussian profile.

\begin{figure}
\includegraphics{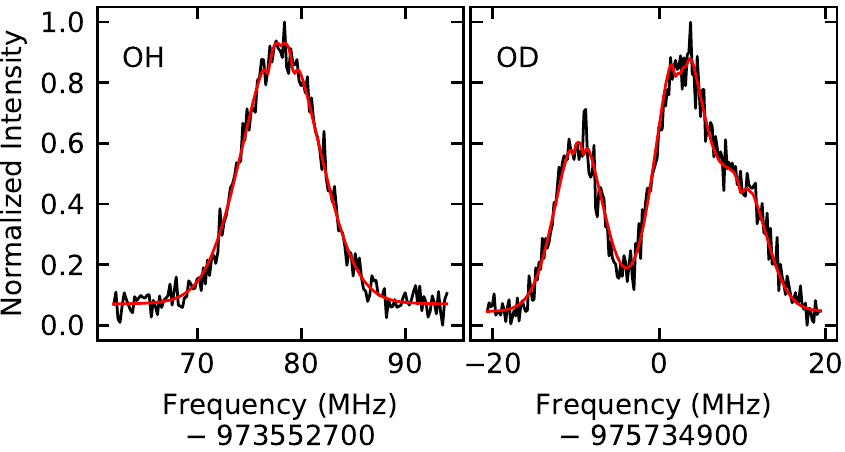}
\caption{Typical relative fluorescence intensity curves (in black) for the $N'=1,J'=1/2,F'=0 \leftarrow f,F''=1$ transition in OH (transition \#4 in table \ref{tab-OHtrans}) and a cluster of transitions to the $N'=1,J'=3/2,F'=3/2$ state from $f,F''=1/2$, $3/2$, and $5/2$ (including transition \#6 in table \ref{tab-ODtrans}).  The fluorescence intensities are extracted by subtracting the background components of the fit shown in figure \ref{fig-osci} and integrating over a $\pm 2$ standard devation region around the fluorescence peak.  Red overlayed curves show the results of the full quantum-mechanical fits described in section \ref{sec-sys-ac}.}
\label{fig-ohod}
\end{figure}
Figure \ref{fig-ohod} shows typical fluorescence intensity versus laser frequency curves for a single transition in OH and a single transition cluster in OD.
Measured curves are shown in black while simulated curves from a detailed fit described in section \ref{sec-sys-ac} are shown in red.  
For OH, all hyperfine transitions were separated by much more than the measured linewidth, so a single transition to each accessible $A$-state level, generally the strongest transition from the $J''=3/2,\Omega''=3/2$ ground state, was measured.
With OD, however, most transitions were blended, due to a smaller ground-state hyperfine splitting, and it was necessary to scan over all allowed transitions from the multiple ground-state hyperfine levels to each excited-state hyperfine level.
Fortunately, the ground-state splittings are quite well known (see section \ref{sec-anal-ham}), which simplified the subsequent analysis.
Each transition or transition cluster was measured at least twice, on separate days, to ensure the measurements are reproducible.   

\section{Systematic effects}\label{sec-sys}
In order to extract the line positions from these measurements, we must first account for possible systematic shifts to the observed line positions.
In this section, we discuss what we consider to be the three largest systematic error sources in this experiment: the quality of the retroreflection of the spectroscopy laser, Zeeman shifts, and the combination of ac Stark shifts and saturation effects.
While OH and OD are both equally affected by retroreflection quality, the effects of Zeeman and ac Stark shifts are larger in OD, due to its smaller hyperfine splittings.
\subsection{Retroreflection quality}\label{sec-sys-ref}
The transitions being measured are subject to Doppler shifts due to the motion of individual molecules along the spectroscopy laser's propagation direction.
While the laser's propagation direction should, in principle, be perpendicular to the molecular beam, an error as small as \SI{100}{\micro\radian} (\SI{0.006}{\degree}) would result in a shift of the measured transition frequency of \SI{110}{kHz}.  
Such systematic Doppler shifts can be largely mitigated by retroreflecting the spectroscopy laser.
For perfect retroreflection (i.e.\ a reflected beam that has the same intensity and profile as the original beam and is exactly anti-parallel), any systematic Doppler shifts for molecules excited by the outgoing beam will have the opposite sign in molecules excited by the retroreflected beam.
If the reflected beam is not exactly anti-parallel, however, a small residual Doppler shift can remain.  

As stated in section \ref{sec-setup-mol}, a \SIadj{0.8}{mm} aperture was placed near the laser source, and the pointing of the retroreflected beam was adjusted for maximum retransmission through the aperture.  
Including the effect of the \SIadj{2}{m} collimation lens, an offset between the outgoing and returning beams at this aperture of \SI{0.1}{mm} would correspond to an angle of \SI{40}{\micro\radian} between the two beams in the spectroscopy region, leading to a \SIadj{\sim 23}{kHz} shift of the measured transition frequency if the offset is along the direction of the molecular beam. 
Based on the sensitivity of the retroreflected signal to small transverse adjustments of the aperture position, as well as the precision with which we can reproducibly adjust the pointing of the mirrors along the beamline, the \SI{0.1}{mm} and \SI{40}{\micro\radian} values were found to be reasonable estimates of the reproducibility of the retroreflection.  
Over the course of each measurement, the pointing was frequently reoptimized, leading to the magnitude and sign of this residual error to be continually randomized.
Additionally, only offsets along the molecular beam direction result in a shift.
Based on this, we estimate the overall contribution of pointing differences between the outgoing and returning beams to the uncertainty of each measurement of a transition frequency to be less than \SI{10}{kHz}.

A mismatch between the intensities of the outgoing and retroreflected beams can also potentially induce a shift in the measured transition frequencies.  
The returning beam must necessarily have a lower intensity than the outgoing beam due to losses in the retroreflection mirror and Brewster window.
If the returning beam excites a different velocity class from the outgoing beam, the stronger weighting of excitation by the outgoing beam will result in a net shift.
The single-pass transmission through the Brewster window has been measured to be \SI{98.5}{\percent}, and the retroreflection mirror in our setup has a UV-enhanced aluminum coating with a reflectance of \SI{93}{\percent} at \SI{308}{nm}.
As a result, the returning beam only has \SI{90}{\percent} of the intensity of the outgoing beam in the interaction region.
To mitigate the influence of this effect, we measured select transitions both with and without the retroreflection beam and adjusted the angle of the spectroscopy laser such that no significant difference could be observed in the transition frequencies.
Comparisons between the spectra with and without the retroreflected beam show no detectable difference in the line shape once this alignment step has been performed, indicating that there is no significant asymmetry in the molecular velocity distribution along the laser axis.
After performing this alignment step, we estimate that the additional error contributed by the amplitude mismatch is less than \SI{5}{kHz} and therefore assign an overall uncertainty to the transition frequency due to retroreflection quality of \SI{10}{kHz}.

While the random fluctuations of the beam pointing can result in an uncorrelated \SIadj{10}{kHz} error for each transition frequency measurement, it is also possible that, due to deviations from a TEM$_{00}$ Gaussian beam profile, maximum retransmission of the returning beam through the aperture does not exactly correspond to perfect retroreflection.  
Measurements of the laser beam profile in the direction of the molecular beam using a knife edge indicate that approximately \SI{80}{\percent} of the power is in $\rm{TEM}_{0n}$ Hermite-Gaussian modes, and the remaining \SI{20}{\percent} of the power is in $\rm{TEM}_{1n}$ modes.
Combining this with the observation that approximately one half of the returning beam passes through the \SI{0.8}{mm} aperture, we conclude that there could potentially be an offset of as much as \SI{0.27}{mm} between the intensity maximum and the true beam center, corresponding to a \SIadj{60}{kHz} offset of the measured transition frequencies.
Since there were no significant changes to the beam path during the measurements, this offset is expected to be the same for all measurements.

\subsection{Zeeman shift}\label{sec-sys-zeeman}
All measurements were conducted in the ambient magnetic field present in the laboratory, with no active compensation.
Using Hall-effect probes (HMMT-6J04-VR and HMNA-1904-VR from Lake Shore Cryotronics), the field strength in the spectroscopy region was measured to be \SI{74}{\micro T} in the vertical direction, \SI{14}{\micro T} along the spectroscopy laser's propagation direction, and \SI{2}{\micro T} in the horizontal direction perpendicular to the spectroscopy laser.  
Since the laser's polarization is horizontal, the magnetic field is almost exactly perpendicular to the laser's polarization axis.  
The field measurements were carried out with a vented vacuum chamber through an open CF40 flange.
To determine whether the turbomolecular pumps and other devices that are running during the experiment affect the field, we measured the field at a fixed point just outside the chamber both with the pumps off and the pumps on and found no measurable change.

Since the magnetic field is, to good approximation, perpendicular to the laser polarization, the transitions observed are those  with $\Delta\!M_F = \pm 1$.
In the limit of a small magnetic field, the shifts of the $\Delta\!M_F = +1$ transitions would be equal and opposite to those of the $\Delta\!M_F = -1$ transitions, and the blended line would show no net shift.  
For larger shifts, however, the states can mix with other hyperfine components, resulting both in a deviation from the linear Zeeman shift and a change of the transition strength.
Since one of the states will be shifted closer to the other hyperfine component while the other will be shifted further away, a non-zero net shift can appear.

To account for the effect of the magnetic field, we make use of an effective Hamiltonian model computed by the program PGOPHER \cite{western16} to predict the eigenstates and eigenenergies of a molecule in the field.
Further details of this effective Hamiltonian model are discussed in section \ref{sec-anal-ham}.
The zero-field parameters for the preliminary model were determined by first estimating approximate line positions through Gaussian fits of the measured spectra, ignoring the Zeeman shifts, and then fitting the zero-field parameters to these line positions; the absolute line positions predicted by this fit were within \SI{250}{kHz} of the final line positions.
Zeeman terms were then included in the Hamiltonian assuming a \SIadj{75}{\micro T} magnetic field perpendicular to the laser polarization and magnetic $g$-factors of $g_L = 1$ for the orbital angular momentum in the electronic ground state, $g_S = 2.002$ for the electron spin, and $g_r = 0$ for the rotational angular momentum.
The eigenstates and eigenenergies predicted by this model are then used by the subsequent fits of the spectra described in sections \ref{sec-sys-ac} and \ref{sec-anal-zero}.

\subsection{AC Stark shift and saturation effects}\label{sec-sys-ac}
To estimate the shifts of the extracted transition frequencies caused by the spectroscopy laser's time-varying (``alternating current'' or ac) electric field and distortions of the spectra due to saturation effects, the spectra were simulated by computing the evolution of the quantum-mechanical density matrix described by the Lindblad master equation (see equation \ref{eqn-lindblad}) using the preliminary model described in section \ref{sec-sys-zeeman} as a basis.
Full details of this calculation are discussed in the \hyperref[sec-appendix]{Appendix}.
The simulation is fit to each measured spectrum using five free parameters, $p_0$ -- $p_4$, using the fitting function
\begin{equation}
I(\nu) = p_0+p_1 S(\nu+p_2,p_3,p_4)\rm{,}
\label{eqn-qmmod}
\end{equation}
where $S$ is the simulated spectrum given by the approximate evaluation of equation \ref{eqn-simspec}.  
Figure \ref{fig-ohod} shows two such fits, in red, for a transition in OH and a transition cluster in OD overlayed on the measured traces.
An absolute, zero-field line positions can extracted from each fitted spectrum by subtracting $p_2$ from the zero-field line positions predicted by the preliminary model, while the Gaussian linewidth due to Doppler broadening and the effective laser power as determined by the relative peak heights can be determined by $p_3$ and $p_4$, respectively.  

While the values of $p_3$ and $p_4$ that result from the fit are generally in the right order of magnitude, there are some slight inconsistencies.  
The fitted values for the Gaussian linewidth cluster around $\sigma = \SI{3.3}{MHz}$ for OH and \SI{3.0}{MHz} for OD, with a standard deviation of \SI{0.15}{MHz} for each isotope; no corresponding shift between OH and OD is observed in the forward velocity of the molecular beam based on the arrival time distribution.
For OD, the fitted values for the effective laser power tend to cluster between 0.2 and 0.4 times the measured power, with several outliers on transitions where the peaks are heavily blended and the relative peak heights cannot be determined.
The reason for these discrepancies is not clear and may indicate there are still some effects not captured by this model.  
Because only one peak is observed in each OH spectrum, there is insufficient information to determine the laser power in these fits; for this reason, the laser power in the fits of OH transitions has been constrained to between 0.2 and 0.4 times the measured power.

\section{Analysis}
\subsection{Zero-field line positions and uncertainties}\label{sec-anal-zero}
Each measured spectrum has been fit using the model described in section \ref{sec-sys-ac}, resulting in an estimate of a transition frequency $\nu_i$ and a corresponding statistical uncertainty $\sigma_i$.
To account for fluctuations of the direction of the retroreflected beam, each $\sigma_i$ is augmented with the \SIadj{10}{kHz} uncorrelated uncertainty estimated in section \ref{sec-sys-ref} using a Pythagorean sum, i.e.
\begin{equation}
\sigma_i'^2 = \sigma_i^2 + (\SI{10}{kHz})^2\rm{.}
\end{equation}
Since each transition or transition cluster was measured at least twice, all measurements of the same transition have been combined in a weighted mean with $\sigma_i'^{-2}$ as the weighting factor.
The overall uncertainty of a transition frequency can be estimated either by combining the individual $\sigma_i'$ or by computing the standard deviation of the (unweighted) mean of the individual measurements; the errors reported in this work represent the larger of the errors determined by these two methods for each transition:
\begin{equation}
\sigma_{\rm{tot}} = \max \Biggl(\Bigl(\sum \sigma_i'^{-2}\Bigr)^{-1/2}, \sqrt{\frac{\sum \bigl(\nu_i - \sum \nu_i/N\bigr)^2}{N(N-1)}}\Biggr)\rm{.}
\end{equation}
Here, the sums are computed over all measurements of a particular transition, and $N$ is the number of measurements of that transition.

To evaluate the influence of the systematic effects described in sections \ref{sec-sys-zeeman} and \ref{sec-sys-ac} on the extracted line positions, the theoretical profiles determined by fitting each experimental spectra with the model described in section \ref{sec-sys-ac} were fit using a simpler model based on a sum of Voigt profiles.
The fitting function for this simplified model is given by
\begin{equation}
I(\nu) = p_0 + p_1 \sum_i p_2 \Bigl(1 - \exp\bigl(-\mu_i^2/p_2\bigr)\Bigr) V(\nu - \nu_i + p_3, p_4, \Gamma)\,
\end{equation}
where $V(\nu, \sigma, \Gamma)$ is a Voigt profile with a Lorentzian width given by $\Gamma$ and a Gaussian width given by $\sigma$, $\nu_i$ and $\mu_i$ are the frequencies and transition dipole moments, respectively, of the individual transitions predicted by the effective Hamiltonian in section \ref{sec-sys-zeeman}, and $p_0$ -- $p_4$ are fitting parameters.
The parameter $p_2$ is an empirical saturation parameter introduced to account for the different relative peak strengths in the OD spectra; for OH, where there is insufficient information to determine the degree of saturation, $p_2$ is assumed to be very large, and the prefactor $p_2 (1 - \exp(-\mu_i^2/p_2))$ is replaced with $\mu_i^2$.
Similar to the previous fit, the line position is determined by subtracting $p_3$ from the zero-field transition frequency predicted by the preliminary model.  
Each theoretical spectrum was modeled using three different values of magnetic field strength (\SI{70}{\micro T}, \SI{75}{\micro T}, and \SI{80}{\micro T}) in the effective Hamiltonian.
Since a field strength of \SI{75}{\micro T} is assumed in the full quantum-mechanical model, we interpret the difference between the simplified model fit using \SI{75}{\micro T} and the full fit to represent the effect of the ac Stark shift and saturation effects.
Similarly, we interpret the difference between \SIadj{75}{\micro T} simplified model fit and the fits assuming \SI{70}{\micro T} or \SI{80}{\micro T} as the effect of a \SIadj{5}{\micro T} variation of the magnetic field strength.
Based on the large discrepancy between the measured power and the power determined based on the fit of the OD spectra, we conservatively assign an error due to the ac Stark shift equal to the total value of the shift.
For the Zeeman shifts, we estimate that the measurement error for the magnetic field strength could be as large as $\pm$\SI{5}{\micro T} and thus estimate a shift due to the magnetic field strength uncertainty of half the difference between the positions determined with the \SI{70}{\micro T} and \SI{80}{\micro T} simplified model fits.

It is not expected that the nature of the power discrepancy or the absolute magnetic field strength will change over time, so these errors are assumed to produce correlated shifts of the transitions.
Because of this, these errors, as well as the \SIadj{60}{kHz} correlated shift described in section \ref{sec-sys-ref}, are not included in the statistical errors, but are considered when determining the errors of the parameters resulting from effective Hamiltonian fit described in section \ref{sec-anal-ham}.
To estimate the overall shift of a transition due to the correlated uncertainties, the shifts are computed for each spectrum individually and combined using a weighted mean, with $\sigma_i^{-2}$ as the weighting factor.
Table \ref{tab-sys} provides a summary of the correlated and uncorrelated shifts due to the systematic effects described in section \ref{sec-sys}.

\begin{table}
\caption{Summary of the transition frequency uncertainties resulting from systematic effects.}
\begin{tabular}{p{28mm}|p{54mm}}
\hline
\hline
Retroreflection \newline quality&	Correlated: \SI{60}{kHz} \newline Uncorrelated: \SI{10}{kHz}\\
\hline
Zeeman shift&	Correlated: $(\nu_{\rm{simp},\SI{80}{\micro T}} - \nu_{\rm{simp},\SI{70}{\micro T}})/2$\\
\hline
AC Stark shift and \newline saturation effects&	Correlated: $\nu_{\rm{QM},\SI{75}{\micro T}} - \nu_{\rm{simp},\SI{75}{\micro T}}$\\
\hline
\hline
\end{tabular}
\label{tab-sys}
\end{table}

The zero-field transition frequencies, together with the corresponding uncertainties, are summarized in table \ref{tab-OHtrans} for OH and table \ref{tab-ODtrans} for OD.
All measured transitions originate from the $X\, ^2\Pi_{3/2}, v''=0,J''=3/2$ rovibronic ground state.
The $p''$ and $F''$ columns indicate the ground-state parity and $F$ quantum number, respectively, while the $N'$, $J'$, and $F'$ columns indicate the excited-state ($A\, ^2\Sigma^+,v'=0$) quantum numbers.
One standard deviation statistical uncertainties are indicated, in units of the last digit, in parentheses next to the frequency, and deviations between the observed and calculated (see section \ref{sec-anal-ham}) transition frequencies are shown in the column labelled ``O$-$C''.  
For OD, the effect of the uncertainties of the magnetic field strength and optical field strength on each transition frequency are shown in table \ref{tab-ODtrans} in the columns labelled ``Zee'' and ``AC'', respectively; for OH, such effects were determined to be negligibly small ($< \SI{2}{kHz}$) and are thus omitted from table \ref{tab-OHtrans}.
Although not shown in table \ref{tab-ODtrans}, the effect of neglecting magnetic fields in the model entirely was also computed and generally found to be about 7 times larger than values shown in the ``Zee'' column, with a deviation as high as \SI{300}{kHz} in the case of transition \#5.
Due to multiple blended transitions being fit simultaneously, all transitions in OD are shown as if they originated from an $F''=1/2$ level, even if this transition would be forbidden by angular momentum selection rules.
Transition frequencies have been corrected for the recoil shift by subtracting $h \nu^2 / (2 m c^2)$, where $h$ is Planck's constant, $\nu$ is measured transition frequency, $m$ is the mass of the molecule, and $c$ is the speed of light.  
These corrections amount to approximately \SI{124}{kHz} for OH and \SI{118}{kHz} for OD.

\begin{table}
\caption{Measured transition frequencies for OH.}
\begin{tabular}{r|ll|lll|l|r}
\hline
\hline
\#&	$p''$&	$F''$&	$N'$&	$J'$&	$F'$&	\mc1{c|}{Frequency [MHz]}&	\mc1c{O$-$C [kHz]}\\
\hline
1&	$e$&	1&	0&	1/2&	0&	\num{972543544.417}(26)&	\num{-3}\\
2&	$e$&	2&	0&	1/2&	1&	\num{972544263.293}(20)&	\num{1}\\
3&	$f$&	2&	1&	1/2&	1&	\num{973552522.962}(27)&	\num{10}\\
4&	$f$&	1&	1&	1/2&	0&	\num{973552777.917}(27)&	\num{9}\\
5&	$f$&	1&	1&	3/2&	1&	\num{973562502.848}(38)&	\num{-15}\\
6&	$f$&	2&	1&	3/2&	2&	\num{973562933.668}(41)&	\num{-25}\\
7&	$e$&	2&	2&	3/2&	2&	\num{975583190.355}(38)&	\num{-21}\\
8&	$e$&	1&	2&	3/2&	1&	\num{975583518.439}(21)&	\num{-13}\\
9&	$e$&	1&	2&	5/2&	2&	\num{975600025.186}(28)&	\num{17}\\
10&	$e$&	2&	2&	5/2&	3&	\num{975600407.540}(27)&	\num{16}\\
11&	$f$&	2&	3&	5/2&	3&	\num{978623067.641}(37)&	\num{-8}\\
12&	$f$&	1&	3&	5/2&	2&	\num{978623423.479}(78)&	\num{41}\\
\hline
\hline
\end{tabular}
\label{tab-OHtrans}
\end{table}

\begin{table}
\caption{Measured transition frequencies for OD.}
\begin{tabular}{r|ll|lll|l|rrr}
\hline
\hline
\#&	$p''$&	$F''$&	$N'$&	$J'$&	$F'$&	\mc1{c|}{Frequency}&	\mc1c{Zee}&	\mc1c{AC}&	\mc1c{O$-$C}\\
&	&		&	&	&		&	\mc1{c|}{[MHz]}&				\mc3c{[kHz]}\\
\hline
1&	$e$&	1/2&	0&	1/2&	1/2&	\num{975191151.074}(33)&	\num{20}&	\num{-42}&	\num{-28}\\
2&	$e$&	1/2&	0&	1/2&	3/2&	\num{975191328.844}(35)&	\num{3}&	\num{1}&	\num{44}\\
3&	$f$&	1/2&	1&	1/2&	3/2&	\num{975729510.621}(13)&	\num{5}&	\num{0}&	\num{-5}\\
4&	$f$&	1/2&	1&	1/2&	1/2&	\num{975729554.271}(68)&	\num{2}&	\num{-23}&	\num{63}\\
5&	$f$&	1/2&	1&	3/2&	1/2&	\num{975734850.974}(54)&	\num{-42}&	\num{-34}&	\num{-21}\\
6&	$f$&	1/2&	1&	3/2&	3/2&	\num{975734909.606}(45)&	\num{5}&	\num{11}&	\num{-5}\\
7&	$f$&	1/2&	1&	3/2&	5/2&	\num{975735003.948}(17)&	\num{3}&	\num{5}&	\num{0}\\
8&	$e$&	1/2&	2&	3/2&	5/2&	\num{976811945.177}(42)&	\num{1}&	\num{7}&	\num{-30}\\
9&	$e$&	1/2&	2&	3/2&	3/2&	\num{976811996.579}(40)&	\num{3}&	\num{-2}&	\num{23}\\
10&	$e$&	1/2&	2&	3/2&	1/2&	\num{976812027.711}(96)&	\num{20}&	\num{-22}&	\num{103}\\
11&	$e$&	1/2&	2&	5/2&	3/2&	\num{976820926.936}(96)&	\num{0}&	\num{-7}&	\num{33}\\
12&	$e$&	1/2&	2&	5/2&	5/2&	\num{976820984.166}(27)&	\num{3}&	\num{-7}&	\num{-42}\\
13&	$e$&	1/2&	2&	5/2&	7/2&	\num{976821062.836}(26)&	\num{2}&	\num{1}&	\num{40}\\
14&	$f$&	1/2&	3&	5/2&	7/2&	\num{978434703.880}(28)&	\num{2}&	\num{1}&	\num{11}\\
15&	$f$&	1/2&	3&	5/2&	5/2&	\num{978434756.718}(37)&	\num{1}&	\num{-12}&	\num{-13}\\
16&	$f$&	1/2&	3&	5/2&	3/2&	\num{978434794.681}(86)&	\num{-28}&	\num{-34}&	\num{-1}\\
\hline
\hline
\end{tabular}
\label{tab-ODtrans}
\end{table}

\subsection{Effective Hamiltonian fit}\label{sec-anal-ham}
The measured transitions were fit to an effective molecular Hamiltonian by varying the $A$-state parameters to minimize the root mean square residuals.  
The effective Hamiltonian used in this work follows, where possible, the linear molecule terms of the 1994 IUPAC recommendation for fine and hyperfine structure parameters \cite{hirota94}, and the quantum numbers follow the 1997 IUPAC recommendation on notations and conventions in molecular spectroscopy \cite{schutte97}.
The rotational part of the Hamiltonian (which is not defined in the IUPAC recommendations) follows the $\hat{N}^2$ convention, i.e.
\begin{equation}
\hat{H}_{rot} = B \hat{N}^2 - D \hat{N}^4 + H \hat{N}^6 + L \hat{N}^8 + M \hat{N}^{10} + P \hat{N}^{12} + Q \hat{N}^{14}\rm{.}
\end{equation}
Centrifugal distortion terms are similarly described in terms of anti-commutators with powers of $\hat{N}^2$.
For example, the spin-orbit Hamiltonian, including centrifugal distortion terms, is given by
\begin{equation}
\hat{H}_{so} = A [\hat{N}^2, \Lambda \Sigma]_+ + A_D [\hat{N}^4, \Lambda \Sigma]_+ + A_H [\hat{N}^6, \Lambda \Sigma]_+ + \cdots
\end{equation}
The fitting procedures were carried out using PGOPHER \cite{western16}, whose definitions of the spectroscopic constants generally match those described above.

\begin{table}
\caption{Parameters for the $X\, ^2\Pi_{3/2}, v''=0$ ground state, in MHz, derived from the global fit by B.\ J.\ Drouin \cite{drouin13}}
\begin{tabular}{lrr}
\hline
\hline
&						$^{16}$OH&				$^{16}$OD\\
\hline
$B$&					\num{555661.4693}&		\num{296158.6891}\\
$D$&					\num{57.2292883}&		\num{16.14328}\\
$H\times10^3$&			\num{4.2810656}&		\num{0.6400}\\
$L\times10^9$&			\num{-448.6944}&		\num{-35}\\
$M\times10^{12}$&		\num{33.315}&			\num{1.3}\\
$P\times10^{18}$&		\num{-838.20}&			\num{-6.0}\\
$Q\times10^{21}$&		\num{-796.21}&			\num{-9.4}\\
$A$&					\num{-4168708.0644}&	\num{-4167841.97}\\
$A_D$&					\num{-17.8685}&			\num{-9.8676}\\
$A_H\times10^{3}$&		\num{18.631}&			\num{5.23}\\
$\gamma$&				\num{-3488.3181}&		\num{-1858.746}\\
$\gamma_D$&				\num{0.61015}&			\num{0.1714}\\
$\gamma_H\times10^6$&	\num{-73.14}&			\num{-11}\\
$p$&					\num{7053.354621}&		\num{3762.01317}\\
$p_D$&					\num{-1.5510938}&		\num{-0.436101}\\
$p_H\times10^6$&		\num{157.746}&			\num{23.97}\\
$p_L\times10^9$&		\num{-28.57}&			\num{-2.3}\\
$q$&					\num{-1160.1202999}&	\num{-328.052845}\\
$q_D$&					\num{0.44211825}&		\num{0.0660521}\\
$q_H\times10^6$&		\num{-82.4266}&			\num{-6.5218}\\
$q_L\times10^9$&		\num{15.1479}&			\num{0.63}\\
$q_M\times10^{12}$&		\num{-2.52506}&			\num{-0.056}\\
$q_P\times10^{18}$&		\num{332.81}&			\num{3.9}\\
$a$&					\num{86.108353}&		\num{13.30473}\\
$b_F$&					\num{-73.155434}&		\num{-11.17400}\\
$c$&					\num{130.643272}&		\num{20.16923}\\
$d$&					\num{56.683092}&		\num{8.77294}\\
$d_D\times10^3$&		\num{-23.007}&			\num{-1.872}\\
$c_I\times10^3$&		\num{-98.9043}&			\num{-8.047}\\
$c_I'\times10^3$&		\num{6.837}&			\num{0.56}\\
$e Q q_0$&				&						\num{0.28569}\\
$e Q q_2$&				&						\num{-0.1205}\\
\hline
\hline
\end{tabular}
\label{tab-gndstatetable}
\end{table}

To model the structure of the $X\, ^2\Pi_{3/2}, v''=0$ ground state, we rely on the comprehensive global fit carried out by B.\ J.\ Drouin \cite{drouin13}, which distills the large body of microwave, pure rotational, and rovibrational spectroscopy that has been carried out on the various isotopologues of OH into a single set of Dunham parameters.  
For our analysis, it was more convenient to convert the Dunham parameters to two separate parameter sets that describe a single vibronic state of each isotopologue, as summarized in table \ref{tab-gndstatetable}.
All transitions measured originate from the various $\Lambda$-doublet and hyperfine components of the $\Omega'' = 3/2, J'' = 3/2$ rotational state.  
For OH, the model yields a term value of the absolute ground state ($e$, $F'' = 1$) of \SI{-589594.229}{MHz}, with the other levels \SI{53.171}{MHz} ($e$, $F'' = 2$), \SI{1665.402}{MHz} ($f$, $F'' = 1$), and \SI{1720.530}{MHz} ($f$, $F'' = 2$) higher in energy.
For OD, the term value of the absolute ground state ($e$, $F'' = 1/2$) is \SI{-1250744.792}{MHz}, with the other levels \SI{7.112}{MHz} ($e$, $F'' = 3/2$), \SI{19.229}{MHz} ($e$, $F'' = 5/2$), \SI{310.143}{MHz} ($f$, $F'' = 1/2$), \SI{317.326}{MHz} ($f$, $F'' = 3/2$), and \SI{329.591}{MHz} ($f$, $F'' = 5/2$) higher.

The previous best values of the $A-X$ transition frequencies in OH and OD come from Stark et al.\ \cite{stark94}, as well as from two earlier papers by Coxon \cite{coxon80,coxon75}.
Additionally, for OH, ter Meulen et al.\ \cite{meulen86} have measured several of the $A$-state spin-rotation splittings with high precision using microwave double-resonance spectroscopy, and for OD, numerous authors have measured the $A$-state hyperfine splittings \cite{german76,carter96,xin03}.
Since we have only been able to measure transitions to the lowest rotational states, we have augmented our effective Hamiltonian fits with information from these previous works, which include transitions to higher rotational levels.
In particular, we have fixed the values for $H$, $L$\footnote{Note that Stark et al.\ use the opposite sign convention for the $\hat{N}^8$ term, so the sign of $L$ has been changed in our parameter list}, $M$, and $\gamma_H$ to those given by Coxon \cite{coxon80} for OH and those given by Stark et al.\ \cite{stark94} for OD.
For OH, we have combined our 12 measurements with the 8 given by ter Meulen et al.\ \cite{meulen86} into a single global fit, and for OD, we have included the $A$-state hyperfine splittings measured by Carter et al.\ \cite{carter96}, as well as those measured by Xin et al.\ \cite{xin03}.

While all OH transitions measured in this work (as well as all transitions from ter Meulen et al.) were included in the effective Hamiltonian fit, two OD splittings, $N=3$,$J=7/2$,$F'=9/2 \leftrightarrow F''=7/2$ from Carter et al. and $N=1$,$J=3/2$,$F'=3/2 \leftrightarrow F''=1/2$ from Xin et al., were excluded as outliers (residuals of more than $3 \sigma$).
There is no indication that any of these anomalies are the result of a deviation between the actual splittings and the effective Hamiltonian model, since the splittings in other data sets that involve the same levels as those in the anomalous splittings do not show any significant deviations from the values predicted by the effective Hamiltonian fit.  
The levels involved in the outlier from Carter et al., however, are not sampled in any of the other data sets.

The effect of the correlated systematic errors on the OD parameters were determined by shifting each measured transition frequency by the offsets given in the ``Zee'' or ``AC'' columns of Table \ref{tab-ODtrans} and refitting the spectroscopic parameters.
The difference between the parameters computed with and without offset is taken as the error due to that systematic effect.
For both OH and OD, an additional systematic error of \SI{60}{kHz} was assigned to the band origin, corresponding to the correlated uncertainty due to retroreflection quality.
All three errors for each parameter are combined into a total systematic error using a Pythagorean sum.

\begin{table}
\caption{Fitted parameters for the $A\, ^2\Sigma^+, v'=0$ state of $^{16}$OH, in MHz.}
\begin{tabular}{lrrr}
\hline
\hline
&						{\centering This work}&			Stark et al. \cite{stark94}& 	Coxon \cite{coxon80}\\
\hline
\multirow{2}{*}{$T$}&		
						\num{971954529.223}&	
													\multirow{2}{*}{\num{971954376}(3)}&		
																						\num{971954664}(54)\\
&						(11)(60)&								&						\num{971954520}(60)\\
$B$&					\num{508601.5809}(53)&			\num{508603.268}(66)&			\num{508599}(2)\\
$D$&					\num{61.87624}(52)&				\num{61.8903}(36)&				\num{61.853}(15)\\
$H\times10^3$&			\num{3.69}\footnotemark[1]&		\num{3.82}(11)&					\num{3.687}(44)\\
$L\times10^6$&			\num{-0.41}\footnotemark[1]&	\num{-0.60}&					\num{-0.412}(57)\\
$M\times10^9$&			\num{-0.11}\footnotemark[1]&	\num{-0.021}&					\num{-0.109}(27)\\
$\gamma$&				\num{6777.832}(10)&				\num{6775.74}(18)&				\num{6762}(10)\\
$\gamma_D$&				\num{-1.43517}(89)&				\num{-1.379}(11)&				\num{-1.430}(54)\\
$\gamma_H\times10^3$&	\num{0.23}\footnotemark[1]&		\num{0.069}&					\num{0.228}(60)\\
$b_F$&					\num{772.077}(26)&				&								\\
$c$&					\num{161.732}(68)&				&								\\
$c_I$&					\num{-0.0335}(77)&				&								\\
\hline
\hline
\end{tabular}
\footnotetext[1]{Parameter held fixed at value from Coxon \cite{coxon80}}
\label{tab-OHparametertable}
\end{table}

\begin{table}
\caption{Fitted parameters for the $A\, ^2\Sigma^+, v'=0$ state of $^{16}$OD, in MHz.}
\begin{tabular}{lrrr}
\hline
\hline
&						This work&						Stark et al. \cite{stark94}&	Coxon \cite{coxon75}\\
\hline
\multirow{2}{*}{$T$}&	\num{973940524.775}&			\multirow{2}{*}{\num{973940470}(3)}&	\multirow{2}{*}{\num{973940860}(60)}\\
&						(21)(62)&						&								\\
$B$&					\num{271124.841}(13)(7)&		\num{271123.980}(48)&			\num{271117.4}(36)\\
$D$&					\num{17.3464}(14)(7)&			\num{17.3428}(13)&				\num{17.2758}(78)\\
$H\times10^3$&			\num{0.56}\footnotemark[1]&		\num{0.561}(36)&				\num{0.4932}(84)\\
$L\times10^9$&			\num{-39}\footnotemark[1]&		\num{-39}&						\num{-18.0}(42)\\
$M\times10^{12}$&		\num{-2.1}\footnotemark[1]&		\num{-2.1}&						\num{-3.51}(87)\footnotemark[3]\\
$\gamma$&				\num{3614.148}(25)(10)&			\num{3616.72}(14)&				\num{3600.6}(63)\\
$\gamma_D$&				\num{-0.4093}(52)(22)&			\num{-0.4011}(66)&				\num{-0.3580}(72)\\
$\gamma_H\times10^6$&	\num{13}\footnotemark[1]&		\num{13}&						\\
$b_F$&					\num{118.468}(20)(18)&			&								\\
$c$&					\num{24.863}(56)(45)&			&								\\
$c_I\times10^3$&		\num{-2.7}\footnotemark[2]&		&								\\
$e Q q_0$&				\num{0.277}(26)(4)&				&								\\
\hline
\hline
\end{tabular}
\footnotetext[1]{Parameter held fixed at value from Stark et al. \cite{stark94}}
\footnotetext[2]{Value from our OH fit, scaled by reduced mass ratio and proton-deuteron g-factor ratio (in total, approximately 0.08136)}
\footnotetext[3]{Represents difference between $M$ in excited and ground states}
\label{tab-ODparametertable}
\end{table}

Tables \ref{tab-OHparametertable} and \ref{tab-ODparametertable} summarize the $A\, ^2\Sigma^+, v'=0$ parameters for OH and OD, respectively, calculated in this work and those from other works based on measurements of electronic transitions.  
One standard deviation statistical uncertainties and uncertainties due to the systematic shifts are indicated, in units of the last digit, in parentheses next to the values, where applicable.  
Because the previous articles use slightly different definitions of the effective Hamiltonian, their parameters have been adjusted to match our definition.
For the $A$ state, the other articles define the centrifugal distortion constants for $\gamma$ in terms of $\hat{J}^2$ instead of $\hat{N}^2$.
Accounting for this requires modifying the parameters as follows:
\begin{align}
B &= B_{\rm{prev}} + \gamma_{D,\rm{prev}}/2 + \gamma_{H,\rm{prev}}/4\\
D &= D_{\rm{prev}} - \gamma_{H,\rm{prev}}\\
\gamma &= \gamma_{\rm{prev}} - \gamma_{D,\rm{prev}}/4 + \gamma_{H,\rm{prev}}/16\\
\gamma_D &= \gamma_{D,\rm{prev}} + \gamma_{H,\rm{prev}}/2
\end{align}
In order to compare the values given for the $A\, ^2\Sigma^+, v'=0$ band origin, the ground state effective Hamiltonian must be considered as well.  
Stark et al.\ \cite{stark94} and the OD paper from Coxon \cite{coxon75} use a ground state Hamiltonian based on a Van Vleck transformation of a Hamiltonian containing a unique perturbing $^2\Sigma^+$ state (for which matrix elements are given in \cite{rao93}), while the OH paper from Coxon \cite{coxon80} uses a Hamiltonian derived using spherical tensor methods, with an $\hat{R}^2$ rotational Hamiltonian \cite{brown78}.
With the Hamiltonians used by Stark et al.\ and in Coxon's OH paper, the $A$-state band origin can be approximately converted to our notation using the formula
\begin{equation}
T' = T'_{\rm{prev}} + B''_{\rm{prev}} + D''_{\rm{prev}}\rm{.}
\label{eqn-conv1}
\end{equation}
The Hamiltonian used in Coxon's OD paper includes an additional $\Lambda$-doubling $o$ parameter, so the band origin from this work has been converted \cite{brown79} using the formula
\begin{equation}
T' = T'_{\rm{prev}} + B''_{\rm{prev}} + D''_{\rm{prev}} - o''_{\rm{prev}}/2\rm{.}
\label{eqn-conv2}
\end{equation}
In equations \ref{eqn-conv1} and \ref{eqn-conv2}, parameters with a single prime refer to $A$-state parameters, while those with a double prime refer to $X$-state parameters.

Based solely on the uncertainties quoted in the previous works, our fits have determined the $A$-state band origins with approximately two orders of magnitude higher precision and the rotational constant $B$ with approximately one order of magnitude higher precision.  
It should be noted, however, that the band origins and rotational constants given by these previous works often differ from our values by much more than the stated uncertainties should allow.  
In the case of Stark et al.\ \cite{stark94}, the band origins for OH and OD differ from our values by $51 \sigma$ and $18 \sigma$, respectively, while the rotational constants $B$ differ by $26 \sigma$ and $18 \sigma$, respectively.
While there is some ambiguity in the conversion of the $A$-state band origins, using another method for the conversion (based on the difference of the term values of the absolute ground state in the two models, ignoring hyperfine effects) only resulted in larger discrepancies.  
While it is difficult to say with certainty why the deviation is so large, we suspect some combination of a pressure shift due to the \SIadj{2.2}{Torr} of helium in the discharge source used in the previous work, as well as errors in the absolute positions of the Fe I lines used for calibration \cite{learner88}.

The parameters determined in the previous hyperfine-resolved studies \cite{meulen86,carter96,xin03} generally show much better agreement with those presented here.
The $b_F$, $c$ and $\gamma$ parameters in ter Meulen et al.\ \cite{meulen86} agree with our parameters for OH to within 2.2 standard deviations; while $\gamma_D$ differs by $4 \sigma$, this discrepancy could be due to the inclusion of a $\gamma_H$ parameter in this work, which was absent in the previous source.
The $b_F$, $c$, and $e Q q_0$ parameters for OD reported by Carter et al.\ \cite{carter96} agree to within two standard deviations, while those in Xin et al.\ \cite{xin03} differ by less than $1 \sigma$. 
A high level of agreement should not be surprising, since the data from all of these works has been included in our parameter fits.
For the same reason, the values reported here also have a slightly smaller uncertainty than those reported previously.  

\section{Conclusions}
The results presented in this work represent some of the most precise Doppler-broadened measurements of molecular electronic transitions to date, comparable in relative precision to the recent measurements of the P7 P7 $B$-band transition in O$_2$ by Bielska and coworkers \cite{bielska17}.
By measuring all experimentally-accessible transitions from the rovibronic ground state, we have also been able to determine the $A$-state spectroscopic constants, particularly the band origin and rotational constants, with far higher precision than in previous measurements.
Fitting multiple measured transitions to an effective Hamiltonian model also serves as a cross-check of the measurement precision of the individual transition frequencies.

We expect that the experimental precision can be improved by an additional order of magnitude by canceling the residual magnetic fields and measuring the transition frequencies with Doppler-free saturation spectroscopy.
Future work will also focus on measuring transitions to higher vibrational levels of the $A$ state.
With sufficient data, we can hopefully start to construct a global model of the $A\,^2\Sigma^+$ electronic state, similar to the one produced by Drouin for the $X\,^2\Pi$ ground state.

\begin{acknowledgments}
We would like to thank Gerard Meijer for his insightful comments regarding the experimental design, Colin M. Western for helpful discussions regarding the use of PGOPHER and its extension with additional spectroscopic parameters, and Alec M. Wodtke and his group for helpful advice and experimental support.
\end{acknowledgments}

\appendix*
\section{Quantum-mechanical simulation of the measured spectra}\label{sec-appendix}
The spectra measured in this work were simulated numerically by computing the evolution of the quantum-mechanical density matrix using the Lindblad master equation \cite{breuer07}:
\begin{equation}
\frac{\partial \rho}{\partial t} = -\frac{i}{\hbar} [H, \rho] + \sum_i \gamma_i \Bigl(A_i \rho A_i^\dag - \frac{1}{2} (A_i^\dag A_i \rho + \rho A_i^\dag A_i)\Bigr)
\label{eqn-lindblad}
\end{equation}
Here, $\rho$ is the density matrix, $H$ is the Hamiltonian, $A_i$ are normalized ``jump'' operators that each describe a spontaneous decay process by transforming a pre-decay state to a post-decay state, and $\gamma_i$ are the corresponding decay rates.  
Equation \ref{eqn-lindblad} is solved in this work using the open-source Python framework QuTiP \cite{johansson13}.

The basis set used in the simulation are the eigenstates from the effective Hamiltonian model described in section \ref{sec-sys-zeeman} (which already includes the effect of the \SIadj{75}{\micro T} static magnetic field).
Eigenenergies in zero optical field and electric transition dipole moments between the individual $X$ and $A$ eigenstates are calculated using PGOPHER.
In order to determine the values of the transition dipole moments in absolute terms, PGOPHER requires a single ``Strength'' parameter, which corresponds to the $A-X$ transition dipole moment in the molecule-fixed frame.  
Based on the \SIadj{688}{ns} lifetime for the $v' = 0, N' = 0$ excited state measured by German \cite{german75}, the value of this parameter is found to be
\begin{equation}
|\langle A, v'=0| T^1_{q=\pm 1}(\vec{\mu}) | X, v''=0\rangle|^2 = 0.26\, \rm{D}\rm{.}
\end{equation}
The transition dipole moments are also used to compute the $M$ state resolved Einstein $A$ coefficients, which are equivalent to the $\gamma_i$ coefficients in equation \ref{eqn-lindblad}.
When simulating the spectrum for a given transition or transition cluster, all $X$ and $A$ levels that are involved in a transition from a $X, v'' = 0, \Omega'' = 3/2, J'' = 3/2$ level to an $A, v' = 0$ level that is no more than \SI{2}{GHz} from the measured spectral region are included in the basis set.
Additionally, all $X, v'' = 0$ levels that have an allowed transition to any of the included $A$ levels are added to the basis set.  
To reduce the number of basis states involved, only transitions for which $\Delta J = \pm 1$ or $0$ are considered; any transitions for which this is not true are only allowed by hyperfine coupling or through mixing due to the magnetic field and thus will be extremely weak.  
For OH/OD, the basis contains 8/12 $X, v'' = 0, \Omega'' = 3/2, J'' = 3/2$ initial levels (corresponding to all levels in either the upper or lower $\Lambda$-doublet component), up to 12/18 $A, v' = 0$ levels, and up to 64/96 additional $X, v'' = 0$ levels to which population in the $A$ levels can decay.

The retroreflected spectroscopy laser is assumed to propagate along the $\hat{y}$ direction with a polarization vector in the $\hat{x}$ direction.
If the optical field is approximated as a pair of collimated counterpropagating Gaussian beams with equal intensity in the outgoing and returning beams, the electric field vector as a function of position and time can be described by the equation
\begin{widetext}
\begin{equation}
E_x(x,y,z,t) = 2 \sqrt{\frac{\mu_0 c P_0}{\pi w_0^2}} \exp\biggl(-\frac{x^2+z^2}{w_0^2}\biggr) \cos\biggl(\frac{2\pi \nu y}{c}\biggr) (\exp(2 \pi i \nu t) + \exp(-2 \pi i \nu t))\rm{,}
\label{eqn-ex}
\end{equation}
\end{widetext}
where $\mu_0 c$ is the impedance of free space (approximately \SI{377}{\ohm}), $P_0$ is the laser power in one propagation direction, $w_0 = \SI{0.5}{mm}$ is the $1/e^2$ beam radius, and $\nu$ is the laser frequency.  
This can be incorporated into a total Hamiltonian given as
\begin{equation}
H = H_0 - E_x \mu_x\rm{.}
\end{equation}
In the matrix representation of this Hamiltonian, $H_0$ is a diagonal matrix containing the eigenenergies of the basis states and $\mu_x$ contains the off-diagonal transition dipole moment matrix elements coupling states of $\Delta M_F = \pm 1$.  

Since the presence of the optical frequency $\nu$ and the large $A-X$ splittings in the Hamiltonian would result in high-frequency oscillations in the density matrix which are difficult to handle numerically, it is helpful to remove them by applying a unitary transformation to the density matrix and invoking the rotating-wave approximation.
Time-dependent unitary transformations of the density matrix $\rho$ given by $\rho' = U^\dag \rho U$, where $U$ is a unitary matrix, result in an equivalent Hamiltonian and jump operators that preserve the form of equation \ref{eqn-lindblad}.
These transformed operators are given by 
\begin{align}
H' &= U^\dag H U - i \hbar U^\dag \frac{\partial U}{\partial t}\label{eqn-transham}\\
A_i' &= U^\dag A_i U\label{eqn-transjump}
\end{align}
In the present case, $U$ is chosen to be a diagonal matrix with $\exp(-2 \pi i \nu t)$ in every position corresponding to an $A$-state level and $1$ in every position corresponding to an $X$-state level.
Based on this definition, the second term of equation \ref{eqn-transham} shifts every $A$-state level by $-h \nu$, while the transformation in the first term multiplies every off-diagonal element in $H$ that connects an $A$-state level to a $X$-state level by either $\exp(2 \pi i \nu t)$ or $\exp(-2 \pi i \nu t)$.  
This effectively converts one of the terms in the last factor of equation \ref{eqn-ex} to $1$ while converting the other to $\exp(\pm 4 \pi i \nu t)$.
The rotating-wave approximation is invoked by setting the rapidly-oscillating term to zero, removing the time dependence of the electric field at a fixed point in space and resulting in a Hamiltonian given by
\begin{equation}
H' = H_0' - 2 \mu_x \sqrt{\frac{\mu_0 c P_0}{\pi w_0^2}} \exp\biggl(-\frac{x^2+z^2}{w_0^2}\biggr) \cos\biggl(\frac{2\pi \nu y}{c}\biggr)\rm{,}
\label{eqn-transham2}
\end{equation}
where $H_0'$ is $H_0$ with all $A$-state levels shifted by $-h \nu$, and the symbols in the second term are as defined for equation \ref{eqn-ex}.
The transformed jump operators for transitions from $A$-state levels to $X$-levels have the form
\begin{equation}
A_i' = \exp(-2 \pi i \nu t) A_i\rm{.}
\label{eqn-transjump2}
\end{equation}
Because $A_i$ always appear in conjugate pairs in equation \ref{eqn-lindblad}, the additional phase factor in equation \ref{eqn-transjump2} cancels out.
As a result, the same time-independent $A_i$ jump operators can be used when solving equation \ref{eqn-lindblad} without changing the result.  

Using the Hamiltonian $H'$ from equation \ref{eqn-transham2} together with the Lindblad master equation (equation \ref{eqn-lindblad}), the evolution of the density operator $\rho$ of a molecule passing through the beam can be calculated.  
The initial value assumed for $\rho$ corresponds to an incoherent mixture of all $X, v'' = 0, \Omega'' = 3/2, J'' = 3/2$ levels in the basis set in equal proportions, and the total fluorescence emitted by the molecule as it passes through the beam is used as a proxy for the signal that would be measured in the experiment.  
In the simulation, the total fluorescence is determined by integrating the population in each $A$-state level and computing a weighted sum with the total fluorescence decay rate as the weighting factor.
Each molecule is assumed to transit the beam with a constant velocity vector.
The velocity in $\hat{z}$ direction fixed at $v_z = $~\SI{340}{m/s}, corresponding to the mean forward velocity of the molecular beam.
Motion in the $\hat{x}$ direction is nearly equivalent to motion in the $\hat{z}$ direction (differing only in relation to the laser polarization and magnetic field axes), and the velocity of molecules in the $\hat{x}$ direction is much smaller than $v_z$, so $v_x$ is fixed to zero.
Four other parameters, specifically the laser frequency $\nu$, the laser power $P_0$, the velocity along the laser propagation direction, $v_y$, as well as the $y$ coordinate of the molecule at its closest approach to the center of the beam ($y_0$) are taken as a variables.  
While the $x$ position (which is constant in a single trajectory) can be taken as a variable as well, the signal resulting from trajectories with a non-zero values for the $x$ position are equivalent to replacing $P_0$ with $P_0 \exp(-2 x^2/w_0^2)$.

If $I(\nu, P_0, v_y, y_0)$ represents the fluorescence emitted by a single molecule passing through the beam, then the total fluorescence that would be expected in the experiment at a single laser frequency would be proportional to
\begin{widetext}
\begin{equation}
S(\nu, P_0, \sigma) = \int_0^\infty \int_0^\infty \int_0^{\frac{c}{4 \nu}} \exp\biggl(-\frac{\Delta_\nu^2}{2 \sigma^2}\biggr) I(\nu, P_0 \exp(-2 x^2/w_0^2), c \Delta_\nu / \nu, y_0) dy_0\,d\Delta_\nu\, dx
\label{eqn-simspec}
\end{equation}
\end{widetext}
Here, $\Delta_\nu$ has been defined as $\nu v_y / c$.
Since each evaluation of $I(\nu, P_0 \exp(-2 x^2/w_0^2), c \Delta_\nu / \nu, y_0)$ is computationally expensive, the integral has been approximated as follows: the inner two integrals are evaluated with fixed values of $P_0 \exp(-2 x^2/w_0^2)$ given by $\SI{1}{mW} \cdot 10^{n/5}$, where $n$ is an integer.
The largest value of $n$ is chosen based on the highest laser intensity used when measuring a particular transition/transition cluster, and the smallest value is chosen such that the value of $I(\nu, P_0 \exp(-2 x^2/w_0^2), c \Delta_\nu / \nu, y_0)$ is approximately linear in $P_0 \exp(-2 x^2/w_0^2)$ for all $\nu$ in the spectral region of interest.  
Over the range from 0 to \SI{2}{MHz}, the integral over $\Delta_\nu$ is approximated as a sum of the integrand at \SIadj{50}{kHz} intervals.
For values of $\Delta_\nu$ larger than \SI{400}{kHz}, $I(\nu, P_0 \exp(-2 x^2/w_0^2), c \Delta_\nu / \nu, y_0)$ is approximately independent of $y_0$, so the integral over $y_0$ is only evaluated for $y_0 = 0$; at values of $\Delta_\nu \le \SI{400}{kHz}$, the integral over $y_0$ is approximated by averaging the values at $y_0 = 0$ and $y_0 = c / (4 \nu)$.
For sufficiently large values of $\Delta_\nu$, the molecule only interacts with one of the two counterpropagating beams that contribute to the standing wave, so the portion of the integral over $\Delta_\nu$ for values larger than \SI{2}{MHz} is evaluated, again with \SIadj{50}{kHz} steps, using the approximation
\begin{equation}
\begin{split}
I\bigl(\nu, P_0 &\exp\bigl(-2 x^2/w_0^2\bigr), c \Delta_\nu / \nu, 0\bigr) \\
\approx\, &I\bigl(\nu + \Delta_\nu, P_0 \exp\bigl(-2 x^2/w_0^2\bigr)/2, 0, 0\bigr)\\
+ &I\bigl(\nu - \Delta_\nu, P_0 \exp\bigl(-2 x^2/w_0^2\bigr)/2, 0, 0\bigr)\rm{.}
\end{split}
\end{equation}
The final integral over $x$ is evaluated using a linear interpolated function between the fixed values of $P_0 \exp(-2 x^2/w_0^2)$ calculated in the previous steps.
To handle regions where $P_0 \exp(-2 x^2/w_0^2) < \SI{1}{mW} \cdot 10^{n_{\rm{min}}/5}$, a value of $0$ is added to the the interpolated function where $P_0 \exp(-2 x^2/w_0^2) = 0$.  

\bibliography{bib}
\end{document}